# Unconventional magnetization in the multiphase superconductor $PdBi_2$

Wenjun Kuang[1,#,*], Ziyi Jiang[1], Lewis Powell[1], Sofiia Komrakova[1], Andre K. Geim[1,2], Irina V. Grigorieva[1,2,*]

[1]Department of Physics and Astronomy, University of Manchester, Manchester M13 9PL, UK.

[2]National Graphene Institute, University of Manchester, Manchester M13 9PL, UK.

# Present address: National Innovation Institute of Defense Technology, AMS, Beijing, China.

* Correspondence to: wenjun.kuang@outlook.com; Irina.V.Grigorieva@manchester.ac.uk

Unconventional superconductors have specific signatures in their magnetic properties, such as intrinsic magnetization at interfaces and around defects and fractional and multiquanta vortices, with much attention focused on heavy fermion and high-$T_c$ superconductors. Here, we report the observation of highly anomalous magnetization in β-$PdBi_2$, a layered superconductor previously shown to exhibit a magnetic field-induced transition from s-wave to nodal p-wave superconductivity at a transition field $H^{*}{\sim}0.1$ T ($H_{C1} < H^{*} < H_{C2}$). This transition is driven by the coupling between spin-polarized electronic bands and the in-plane magnetic field. In the unconventional phase (above $H^{*}$) we observe three striking features: strictly linear and non-hysteretic dc magnetization, a sharp drop in the ac susceptibility when the field is applied parallel to the ab-plane, and a pronounced anisotropy in the magnetic response between parallel and perpendicular field orientations. We show that these features are directly correlated with the expected transition to the nodal p-wave state and propose that the unusual magnetization behavior can be explained by a transition from a conventional vortex lattice in the low-field s-wave phase to a domain structure corresponding to spatial phase separation into superconducting (p-wave) and normal domains above $H^{*}$. Our work identifies new experimental signatures of unconventional multiphase superconductivity, offering an insight into the magnetic-field response of nodal states.

## INTRODUCTION

One of the tell-tale characteristics of unconventional superconductivity is an unusual response to the magnetic field, for example, a material remaining in the superconducting state well above the paramagnetic Pauli limit [1,2], re-entrant superconductivity in high magnetic fields [3,4], half-quantum vortices [5,6], magnetic-field induced phase transitions between different superconducting phases [7,8], vortex lattice symmetries deviating from the conventional triangular lattice [9], or spontaneous magnetization in zero applied field [10]. Yet signatures of unconventional superconductivity in magnetization response to the applied field are little known, likely because it is usually dominated by vortex pinning and other flux trapping effects.

Here we study magnetization and surface susceptibility of an unconventional superconductor β-$PdBi_2$. It is believed to be a promising candidate for topological superconductivity [11,12] due to strong spin-orbit coupling and local ('hidden') breaking of the lattice inversion symmetry [13,14]. The latter gives rise to in-plane polarized spin textures not only in the topological surface states, but also in bulk electronic bands [11,12,14], which was recently shown to lead to a phase transition from conventional s-wave superconductivity in zero- and out-of-plane ($H \| \mathrm{c}$) fields to nodal pairing in sufficiently high in-plane ($H \| \mathrm{ab}$) fields [15].

Earlier studies of the response of $PdBi_2$ to magnetic fields focused on the properties of vortices [16-18], magnetic-field-dependent heat capacity [19,20] and superfluid density [21]. Although most of these findings are consistent with conventional type II superconductivity and single-gap s-wave pairing, some features are more difficult to explain, unless one assumes a more complex nature of its superconductivity. Among these is an unusual evolution of the superfluid density, indicative of the presence of two superconducting gaps, one of them nodal [21], and the very low anisotropy of the critical fields $\gamma = H_{c2}^{||ab}/H_{c2}^{||c} \leq 1.25$ [20,21]. The latter is inconsistent with the largely two-dimensional (2D) electronic structure [22] (in most layered materials γ lies in the range 4-8, e.g., $\gamma = 4.8$ for $MgB_2$ [23]; $\gamma = 7.4$ for FeSe [24]; $\gamma > 10$ for organic superconductors [25], etc.) So far there have been no reports of either the field-dependent dc magnetization, or the ac susceptibility of $PdBi_2$.

Here we report an unusual response of the magnetization and surface susceptibility of β-$PdBi_2$ to the magnetic field, particularly for $H$ applied parallel to the crystal's ab-plane. While the low-field dc magnetization, $M(H)$, is typical for a type-II superconductor, above a certain critical field $H^* \approx 0.3H_{c2}$, we observe strictly linear and fully reversible $M(H)$ accompanied by a strong suppression of the magnetic field screening by surface currents. This is inconsistent with simply changes in vortex pinning, which could be caused by, e.g., vortex lattice melting, but consistent with the appearance of normal domains, such as typically observed in type-I superconductors. In perpendicular fields ($H$||c), where STM imaging found a standard triangular vortex lattice for this material [16], we observe a mixed behavior: while $M(H)$ above $H^*$ is also linear and reversible, the susceptibility remains conventional up to $H_{c2}$, corresponding to a strong anisotropy. We propose that the above features of $PdBi_2$ magnetization are a manifestation of a transition at $H^*$ to coexisting superconducting (p-wave) and normal phases separated in space into superconducting and normal domains. Such spatial phase separation is reminiscent of the intermediate state of type-I superconductors, but in $PdBi_2$ it is driven by the field-induced phase transition from s-wave to nodal p-wave superconductivity [15]. Notably, this scenario is qualitatively different from phase coexistence throughout the superconducting volume, e.g., 2D and 3D gaps in $MgB_2$ [26] or coexisting s-wave- and p-wave pairing channels proposed for few-layer $NbSe_2$ [27].

**RESULTS**

DC magnetization and ac susceptibility measurements of β-$PdBi_2$ single crystals were carried out using a commercial SQUID magnetometer MPMS XL7. The starting bulk crystals were grown using a melt-growth method (details in Supplemental Materials) and cleaved using adhesive tape to produce platelet-shape crystals with flat surfaces and thicknesses in the range $d$ = 15 to 140 μm. To probe the anisotropy of magnetization, measurements were done for two orientations of the magnetic field with respect to the crystal's ab plane: parallel ($H$||ab) and perpendicular ($H$||c). Very similar results were found for all samples, not only qualitatively but also quantitatively, with no obvious dependence of the superconducting characteristics on $d$: critical temperature $T_c = 4.5$K, critical magnetic fields at our lowest measurement temperature $T$ =1.8K $H_{c2}^{||ab} = 6.8$ kOe, $H_{c2}^{||c} = 5.46$ kOe, penetration depth $\lambda(0) \approx$ 240 nm, coherence length $\xi_{ab}(0) \approx$22 nm, and Ginzburg-Landau parameter $\kappa \approx$11, see Supplemental Materials for details. Additionally, several thinner crystals ($d \approx 10$ μm) were incorporated into Hall bar-like devices and used for transport measurements (Supplemental Materials). To prevent surface degradation in air, both the as-grown and exfoliated crystals were always handled in the inert atmosphere of an Ar filled glovebox.

Figure 1a shows typical dc magnetization curves, $M(H)$, for the magnetic field applied parallel to the ab plane, $H$||ab. Although overall $M(H)$ seems to describe a type-II superconductor, there are several

unusual features. At low $H^{\parallel ab} < H_{c1}^{\parallel ab}$, we observe full Meissner diamagnetism with $M = -H/4\pi$ until a cusp at $H_p^{\parallel ab}$ (inset in Fig. 1a) indicating entry of vortices, as expected. However, at higher $H \gtrsim 2$ kOe and up to the upper critical field $H_{c2}^{\parallel ab}$, the magnetization curves are anomalously linear and reversible, Fig. 1b. Such linear and reversible dc magnetization in a broad range of magnetic fields is unexpected: type-II superconductors normally display large hysteresis due to vortex pinning [28,29] and $M(H)$ for intermediate flux densities decreases as $M \propto \ln(H_{c2}/B)$ ( $B = H + 4\pi M$ is the magnetic induction) [30]. Also surprising for a layered material, quantitative characteristics of $M(H)$ for the two field orientations are very similar: Not only the upper critical fields $H_{c2}^{\parallel ab}$ and $H_{c2}^{\parallel c}$ differ by just ~20% (Fig. 1c) and the broad region of linear and reversible magnetization is seen for both $H||$ab and $H||$c (Fig. 1c and Fig. 2) but even the absolute values of the bulk magnetic moment are practically identical in all fields above $H_p$ (Fig. 1c). This means that the same amount of flux penetrates the crystals, implying the same screened volume and the same magnetic field penetration depth λ, which is contrary to the conventional behavior of layered superconductors [31].

Furthermore, the transition to the normal state at $H_{c2}$ is sharply defined, with a clear kink where the linear $M(H)$ crosses the horizontal axis ($M = 0$) for both $H||$ab and $H||$c, Fig. 1b,c and Fig. 2. This is in contrast to the classical $H$-dependent magnetization near $H_{c2}$, where the vortices are packed so tightly that the cores fill much of the volume and $M$ is vanishingly small, particularly for relevant Ginzburg-Landau parameters κ ~10. Accordingly, $M(H)$ normally approaches the $M = 0$ axis smoothly, without a kink [30,32]. Only at temperatures close to $T_c$ do we see conventional magnetization curves, with significant hysteresis over most of the magnetic field range, Fig. 1d.

In contrast to the almost isotropic dc magnetization, ac susceptibility is found to be strongly anisotropic, showing a standard behavior in perpendicular fields but starkly different and unusual features of in-plane susceptibility, see Fig. 3a,b and Fig. 4. We recall that, while dc magnetization of a superconductor detects the bulk magnetic response, ac susceptibility at small excitation amplitudes probes a thin surface layer with thickness $x$~λ [32-34]. In a magnetic field parallel to a surface this corresponds to a surface superconducting sheath, where screening currents generated in response to the ac field can screen the whole volume of the superconductor, mimicking the Meissner state up to $H_{c2}$, irrespective whether the applied dc field is fully expelled from the superconductor (below $H_{c1}$), or it penetrates into the bulk in the form of vortices [35-37]. The in-phase part of the susceptibility, χ′, characterizes this screening and the out-of-phase component, χ″, is due to dissipative processes. As seen in Fig. 3b, for our $PdBi_2$ the perpendicular ac field, $h_{ac}^{||c}$, is fully screened by the surface currents (χ′ ≈ $-1/4\pi$) up to $H_{c2}$, as expected. In contrast, in the case of in-plane ac fields, full screening (χ′ = $-1/4\pi$) was only seen in the Meissner state for $H < H_p^{\parallel ab}$. In the mixed state, the ac fields are mostly screened only up to $H^{\parallel ab} \approx 0.1$ kOe $\ll H_{c2}$, Fig. 3a. Increasing the field beyond this value results in a dramatic decrease in |χ′| accompanied by a peak in χ″ , indicating a transition to a new state. After that both $\chi$ components plateau, with χ′ corresponding to just 15-25% of the full diamagnetic screening, depending on the excitation amplitude $h_{ac}$ (top panel of Fig. 3a). At $H^{\parallel ab} \approx H_{c2}^{\parallel ab}$, χ′ jumps from this small but finite diamagnetic response to the normal state value $\chi_N \approx 0$, accompanied by a peak in χ″.

Neither the sharp drop in χ′($H$), nor the peak in χ″ at $H \sim H^*$ can be explained within the conventional vortex scenario, such as being due to vortex motion. In the latter case χ′ and χ″ would depend on the excitation frequency [38], which is not the case in our experiment, where both are frequency independent (Supplemental Materials). Notably, the range of magnetic fields where χ′ shows a plateau accurately matches the range of linear and reversible dc magnetization (bottom panel of Fig. 3a). We identify the field $H^*$ that corresponds to the peak in $\chi''$ (Fig. 3a) as a new critical field for $PdBi_2$ superconductivity.

A pronounced anisotropy is also apparent in the temperature dependence of the ac susceptibility measured at different dc fields, Fig. 3c and d. For $H||c$, $\chi_\perp(T)$ is barely affected by $H < H_{c2}^{||c}$, featuring full diamagnetic screening below $T_c$ at all fields, as expected for conventional superconductors [33-36]. That is, $\chi'_\perp(T)$ tends to -1/4π below $T_c$ at all fields where the material remains superconducting. In contrast, the in-plane susceptibility, $\chi'_{||}(T)$, is dramatically modified by $H$: A conventional behavior is seen below $H_p$ (in the Meissner state) and just above it, but at higher fields screening is progressively suppressed and in the region of linear and reversible dc magnetization, $\chi'_{||}(T)$ saturates at a fraction of the full diamagnetic response,~15% (compare the bottom panel of Fig. 3a and Fig. 3c).

As seen in Fig. 3a, both susceptibility components, $\chi'$ and $\chi''$ are affected by the ac amplitude, $h_{ac}$, as can be expected due to the critical state nature of the surface screening [35,37,38]. As soon as the screening currents exceed the critical value (typically close to the depairing current), the external ac field starts penetrating deeper into the interior of the sample, beyond the surface superconducting layer. This adds contributions from vortices and other bulk non-uniformities and yields an ac amplitude-dependent response [38-40]. For our $PdBi_2$, similar to the susceptibility anisotropy, the magnitude of the effect is starkly different for parallel and perpendicular fields. For $H||c$, increasing $h_{ac}$ rescales the susceptibility but does not change it qualitatively: A sharp decrease in $\chi'$ and a peak in $\chi''$ followed by a plateau are seen for all $h_{ac}$ used in the experiment and the amplitude dependence is relatively weak. In contrast, for the perpendicular field, the effect of increasing $h_{ac}$ is dramatic: The susceptibility remains conventional ($\chi' \approx -1/4\pi$ and $\chi'' \approx 0$ up to $H_{c2}$) only for the smallest values of $h_{ac}$ used in the experiment, $h_{ac} < 0.1$ Oe; at larger $h_{ac}$ the diamagnetic response sharply decreases and starts to resemble in-plane susceptibility, see Fig. S2 in Supplemental Materials.

Figure 4 emphasizes that the above features of $H$-dependent ac susceptibility and its strong anisotropy for the two field orientations are present at all temperatures within the superconducting state, except very close to $T_c$, see below. As to the new transition field $H^*$ (defined as $H$ corresponding to the peak in $\chi''$), it is almost independent of temperature, particularly given the uncertainty due to the weak dependence of the $\chi''$peak position on $h_{ac}$ (see the middle panel of Fig. 3a).

On the basis of the above data we have constructed a phase diagram shown in Fig. 5a, which includes the new critical field $H^*$. The presence of two distinct regions of magnetization behavior between $H_{c1}$ and $H_{c2}$ is clear over the whole $H - T$ space and is consistent with our earlier tunneling spectroscopy experiments on β-$PdBi_2$ [15], where the abrupt change in tunneling characteristics from low to high magnetic fields was shown to correspond to a transition from s-wave to nodal p-wave pairing. To see this transition as a kink in $H_{c2}(T)$, as in Fig. 2d in ref. [15], it is necessary to zoom-in in the part of the phase diagram near $T_c$, corresponding to $H \sim H^*$. To this end we obtained $H_{c2}(T)$ from resistance measurements (details in Supplementary Materials). The results are shown in Fig. 5c. In agreement with the field-induced transition seen in ac susceptibility, we found a weak but nevertheless clear kink in $H_{c2}(T)$ at $H \sim 0.7$ kOe, consistent with the existence of a different superconducting phase in the low $H$ - high $T$ region and reminiscent of the phase diagram found for ~100 nm thick $PdBi_2$ crystals (from the same batch) studied using transport and tunneling spectroscopy in ref. [15]. In fact, careful measurements of $H_{c2}(T)$ from dc magnetization detected a similar kink (Fig. 5b), while the $M(H)$ curves in this $T$ range (close to $T_c$) showed significant hysteresis and non-linear $M$ (Fig. 1d), as expected for a conventional type-II superconductor and contrasting the linear reversible magnetization at lower $T$.

We note that the functional form of $H_{c2}^{||ab}(T)$ in Fig. 5a of the present manuscript is different from $H_{c2}(T)$ for the magnetic field parallel to the ab plane in Fig. 2d of ref. [15]. The difference is due to different dimensionalities of the $PdBi_2$ crystals studied in these two cases. In ref. [15], we studied thin crystals with thickness $d \sim 100$ nm or less. The data in Fig. 2d of ref. [15] are for a 80nm crystal, i.e., such that $d < \lambda$ ($\lambda(0) \approx$240 nm) and comparable to $2\xi(0) \approx 44$ nm. There are no vortices in the crystal in that case, $H_{c2}$ is strongly enhanced due to the reduced diamagnetic energy and follows the 'thin film' $T$ dependence $H_{c2}^{\parallel}(T) = \frac{\sqrt{3}\Phi_0}{\pi \xi_{ab} d}\sqrt{1-\frac{T}{T_c}}$ [30]. In contrast, here we study bulk crystals with $d \gg \lambda$, $H_{c2}$ is orbitally limited and follows the standard WHH $T$ dependence [41]. As to $H_{c2}^{||c}$, its $T$ dependence in the 'thin film' limit (as in ref. [15]) is linear, according to $H_{c2}^{||c} = \Phi_0/2\pi\xi^2$ and $\xi(T) \propto \left(1-\frac{T}{T_c}\right)^{-1/2}$, while for the bulk crystals here it also follows the WHH dependence, similar to the in-plane field.

The transport measurements also showed that the critical currents remain finite in the whole region of magnetic fields up to $H_{c2}$ , seemingly in contradiction to fully reversible magnetization curves. This behavior is seen both for $H$||ab and $H$||c, see Fig. S3 in Supplemental Materials.

## DISCUSSION

The described behavior of dc magnetization is incompatible with a conventional mixed (vortex) state in the bulk of our $PdBi_2$ crystals, while the strong suppression of $\chi'$ by $H||\text{ab} > H^*$ contradicts the expected Meissner-like state up to $H_{c2}$ due to screening by the superconducting surface sheath. Furthermore, the strictly linear $M(H)$ and a sharp kink at $H_{c2}$ are reminiscent of magnetization of type I (rather than type II) superconductors. In principle, reversible magnetization could indicate vortex melting, a new phase of vortex matter well established for high-$T_c$ superconductors [42-45]. However, in these materials, vortex melting is the consequence of large thermal fluctuations that are themselves due to a combination of high $T_c$ and short coherence lengths. It can be quantified by the Ginzburg number $Gi$ that measures the relative size of the zero-$T$ condensation energy within a coherence volume and the thermal energy (critical temperature $T_c$): $Gi = \left(2k_B T_c/H_c^2\, \varepsilon\xi^3\right)^2/2$ [42]; vortex melting requires $Gi = 0.01 \sim 1$. For β-$PdBi_2$ the thermodynamic critical field $H_c = H_{c2}/\sqrt{2}\kappa \approx 600$ Oe; $T_c = 4.5$ K; the anisotropy parameter $\varepsilon \sim 1$ and $\xi \approx 22$ nm, which gives $Gi \sim 10^{-8}$, far too small for thermal fluctuations to play a role. Furthermore, the dissipation peak in $\chi''$ associated with the vortex melting transition [42,44,45] is both temperature- and frequency dependent, while for $PdBi_2$ the $\chi''$ peak at $H^*$ (which signifies the transition to reversible $M(H)$) is practically $T$- independent (Fig. 4a) and does not depend on frequency (Supplemental Materials). Qualitatively, the importance of thermal fluctuations is reflected in the fact that the melting transition is typically observed close to $H_{c2}(T)$. This is also not the case for our $PdBi_2$, where reversible magnetization sets in at $H \sim 0.3\, H_{c2}$. Finally, our critical current measurements (Fig. S3) showed that $J_c$ is finite in intermediate fields where $M(H)$ is strictly reversible (for both field orientations). In a conventional vortex picture, finite $J_c$ should correspond to a hysteresis in $M(H)$ and can be estimated as $J_c \propto \Delta M/d$ ($d$ is the thickness) [29].

Given our previous finding of a magnetic field-induced phase transition from s-wave- to nodal p-wave pairing in this system [15], similar physics can be expected to be responsible for the unusual behavior of magnetization. Indeed, the weak but nevertheless clear kink in $H_{c2}(T)$ seen at ~$0.95T_c$ both in transport and dc magnetization measurements (Fig. 5b,c) is reminiscent of the (more pronounced) kink seen in transport measurements on ~100 nm thick $PdBi_2$ crystals in ref. [15]. One of the consequences of the existence of different superconducting phases in a material is the possibility of spatial phase separation, as discussed extensively in the literature for e.g. domains of opposite chirality in chiral p-wave superconductors [47,48] or as the result of an internal competition between degenerate superconducting states near phase boundaries, e.g., in some high $T_c$ cuprates [49,50] or between A- and B-phase in $UPt_3$ [7,51]. In such cases an external magnetic field can drive a transition of parts of the superconductor to a different phase or even to separation into superconducting and normal domains as found for a high-$T_c$ superconductor $La_{1.94}Sr_{0.06}CuO_{4+y}$ [50].

We use the analogy with phase separation in known multiphase superconductors and propose that all the unusual features of $PdBi_2$ magnetization for $H\|\mathrm{ab} > H^*$ can be explained by spatial phase separation into p-wave superconducting and normal domains. This includes the linear and reversible $M(H)$, its kink at $H_{c2}$, the reduced screening and pronounced anisotropy in ac susceptibility (Figs. 1 to 4), independence of frequency and temperature, $H^* \ll H_{c2}$, and finite $J_c$. Since $\chi'$ represents the volume fraction fully screened by the circulating surface currents, the ac results for $H\|\mathrm{ab}$ (Fig. 3a) imply that this volume fraction is abruptly reduced above $H^*$ and the magnetic field fully penetrates parts of the $PdBi_2$ crystal. In dc magnetization this should appear as linear and reversible $M(H)$ between $H^*$ and $H_{c2}$ as indeed observed. Furthermore, above $H^*$ the dc magnetization curves for $PdBi_2$ look remarkably similar to those of platelet-shaped type-I superconductors in the intermediate state, where normal and superconducting domains coexist [52]. As an example, Fig. S4 in Supplemental Materials shows our $M(H)$ measurements for a Sn cylinder with a demagnetization factor $N \approx 1/2$. On the other hand, the standard type-II behavior of ac susceptibility for the out-of-plane field (Fig. 3b,d) implies that the ac field is completely screened by the surface currents circulating in-plane. This suggests formation of planar superconducting and normal regions, as schematically shown in Fig. 5a. The normal regions will disrupt the screening currents generated by in-plane fields, as these must cross the ab planes, and lead to a strongly reduced $\chi'$, as observed. This explanation is also consistent with the presence of vortices in $PdBi_2$, which for $H\|\mathrm{c}$ has been reported in the literature [16,53] and also follows from the finite critical currents observed here (Fig. S3).

We emphasize that the proposed spatial phase separation is different from a situation where different phases coexist throughout the superconducting volume, giving different contributions to the overall behavior but not being separated in space. This is the case, e.g., for 2D and 3D gaps in $MgB_2$ [26] or s-wave- and p-wave pairing channels in few-layer $NbSe_2$ [27]. In contrast, the proposed p-wave superconducting and normal phases in $PdBi_2$ are separated into different domains.

At first sight it may seem surprising that in the phase diagram (Fig. 5) the transition at $H^*$ from s-wave to p-wave pairing produces only a weak anomaly in $H_{c2}(T)$, weaker even than a similar feature for thinner $PdBi_2$ crystals in ref. [15]. Here it is important to distinguish between features corresponding to a phase transition deep inside the superconducting state (low $T$, $H \ll H_{c2}$) and at the boundary between the superconducting and normal states. Away from the transition to the normal state the anomalies in the field dependence of the superconducting gap in ref. [15] and in the ac susceptibility for $H\|\mathrm{ab}$ in the present work are very clear, Fig. 2 in ref. [15] and Fig. 3a,c here. On the other hand, an anomaly at the superconductor-normal state boundary would only be strong if the two phases have sufficiently different $T_c$ and critical magnetic fields. This appears not to be the case for β-$PdBi_2$: for the thin crystals in ref. [15] we estimated $T_c^{p-\mathrm{wave}}/T_c^{s-\mathrm{wave}} \approx 0.8$ ($T_c^{p-\mathrm{wave}}$ is taken as an extrapolation to $H = 0$ in Fig. 2d) and the underlying $H_{c2}(0)$ for the two $H$ orientations appear to be close in value, as most of the $H_{c2}$ enhancement for $H\|\mathrm{ab}$ can be explained by the 'thin film' limit (see above). For the bulk crystals, the difference in $T_c$ cannot be estimated in a similar way because the anomaly is too close to $T_c(0)$, Fig. 5b,c, and also the transition is not between pure s-wave and pure p-wave as in ref. [15] but between s-wave phase and an intermediate state-like domain structure, which can reasonably be expected to have critical parameters different from the pure p-wave phase. In this respect we note that the value of $H^*$ for the bulk crystals (Fig. 5) is notably lower than in ref. [15] (~0.07 T vs 0.2 T), consistent with the described differences.

The fact that linear and reversible dc magnetization is also observed for $H\|\mathrm{c}$, where a transition to p-wave pairing is not expected [15], is likely to be due to the large aspect ratio of the crystals used in our study: even when $H$ is nominally perpendicular to the ab plane, the large demagnetization factor leads to a sizable in-plane field component that can exceed the relatively low value of $H^*$ and result in a domain structure similar to the case of $H\|\mathrm{ab}$. Indeed, when we increased the amplitude of the ac field used in the susceptibility measurements, $h_{\mathrm{ac}}^{\perp}$, both $\chi'$ and $\chi''$ started to resemble the results for $H\|\mathrm{ab}$ (Fig. S2 in Supplemental Materials). While for the smallest $h_{\mathrm{ac}}^{\perp}$ the generated in-plane supercurrents are confined to the superconducting surface sheath and therefore probe the true, Meissner-like, response to the perpendicular field (Fig. 3b), the increasing depth to which the ac field penetrates at larger $h_{\mathrm{ac}}^{\perp}$ [35,37] starts to probe the interior of the crystal, where a domain structure can be expected as described above.

While the proposed spatial phase separation and domain structure can explain all the details of $PdBi_2$ magnetization, the question of why the superconductor would choose an intermediate, rather than the vortex, state after the phase transition remains open and requires further development of theory. Attractive intervortex interactions and double quanta vortices have been suggested for p-wave superconductors due to the vectorial nature of the order parameter [54-56]. It is then conceivable that the normal domains proposed here may represent an extension to multiquanta vortices. Additionally, as the p-wave phase appears to be favored at the surfaces of $PdBi_2$ [15], the system may choose to create additional 'surfaces' at domain boundaries in order to support it. The proposed structure will then be reminiscent of the intermediate-mixed state observed in pure Nb [57], but with the mixed state (vortex) domains coexisting with the normal - rather than Meissner – phase.

Although most new features of $PdBi_2$ superconductivity are found at $H > H^*$, it is instructive to also discuss its magnetization below the transition to p-wave, particularly the role of surface and geometrical barriers [58-61]. Surface (Bean-Livingston) barriers lead to magnetic irreversibility due to the asymmetry between flux penetration (delayed by the barrier) and its exit [58] and are important for superconductors with smooth flat surfaces in parallel magnetic fields. These conditions are satisfied for our $H\|\mathrm{ab}$ $M(H)$ measurements and a finite barrier for vortex penetration does appear to be present: The penetration field $H_{\mathrm{p}}(0) \approx 90$ Oe (extrapolated from $H_{\mathrm{p}}(1.8\mathrm{K}) = 70$ Oe, inset in Fig. 1a) is notably larger than $H_{\mathrm{c1}}^{\|\mathrm{ab}}(0) \approx 53.5$ Oe reported in the literature [18]. An even clearer fingerprint is a flat $M(H)$ curve, $M$ ≈const, in decreasing $H$ that sets in at $H{\sim}H^*$ (Fig. 1a). Although this constant $M$ in not close to zero, as normally expected for BL barriers [58,59], the finite diamagnetic value is 'inherited' from the intermediate-like p-wave phase above $H^*$ (we recall that $H^*$ is an effective upper critical field for the s-wave phase, as discussed in detail in ref. [15]). Only at $H$ just above $H{\sim}H_{\mathrm{c1}}$ does $M$ drop sharply to zero and the sample returns to the flux-free state at $H = 0$, as expected for weak pinning. Taken together, an enhanced $H_{\mathrm{c1}}^{\|\mathrm{ab}}$ and constant $M$ below $H^*$ indicate the importance of the surface barrier for $PdBi_2$ magnetization in the s-wave phase, as can be expected for an exfoliated layered crystal in $H\|\mathrm{ab}$. Detailed interpretation of the above observations requires understanding of the nature of vortices in the proposed p-wave domains and their evolution into conventional vortices below $H^*$; such understanding is currently lacking.

We also observe main manifestations of geometrical barriers in the low-field region for $H\|\mathrm{c}$ [60,61], Fig. 2: a large hysteresis at low $H$ (much larger than for $H\|\mathrm{ab}$ in the same field range, Fig. 1a), significant asymmetry between positive and negative fields, consistent with simulations for a geometrical barrier combined with $H$-dependent pinning (Fig. 7 in ref. [61]), and a sharp peak near zero applied field. The latter is particularly pronounced in our case because of a large aspect ratio for this crystal, $d/a \approx 0.045$. Using the measured penetration field $H_p = 8$ Oe, we have estimated the lower critical field for $H \| c$ using $H_{c1} = H_p/\tanh\sqrt{0.36d/a}$ [61]. This yielded $H_{c1}^{\|c}(1.8K) \approx 60$ Oe, in good agreement with the estimate obtained from measured penetration depth in ref. [18], $H_{c1}^{\|c}(0.9K) \approx 58$ Oe.

Finally, we note that the surprisingly low anisotropy of the critical fields for $PdBi_2$, $H_{c2}^{||ab}/H_{c2}^{||c} \approx 1.2$, can be explained by the finite dispersion of its quasi-2D Fermi surface in $k_z$ direction [22], similar to some layered iron pnictides [62,63]. These can be sufficient to permit circulating currents and orbital limiting effects to persist for all field directions, leading to near isotropic $H_{c2}$ [62]. An alternative or additional explanation can be that several electronic bands contribute to superconductivity, possibly with opposing anisotropy parameters, leading to an isotropic overall behaviour as reported in ref. [63].

In summary, we report anomalously linear and reversible magnetization over a large part of the $H - T$ phase diagram of the superconducting β-$PdBi_2$, reminiscent of the intermediate state of type-I superconductors. At the same time our ac susceptibility measurements indicate the lack of continuous superconducting currents circulating out of plane (for in-plane magnetic fields) above a certain critical field $H^*$. We propose that these features of magnetization can be explained by the magnetic field induced phase transition to nodal p-wave paring reported for this material in our previous work [15], which in bulk crystals studied here appears to result in spatial phase separation into normal and superconducting domains. We emphasize that the proposed qualitative explanation is only one possibility and other scenarios cannot be excluded. Fully understanding superconductivity of this material, which is a representative of a class of centrosymmetric superconductors with 'hidden' inversion symmetry breaking, requires further development of theory and alternative experimental tools to detect inhomogeneous magnetic fields and superconducting properties at the microscopic scale.

**Acknowledgements.** We acknowledge financial support from Horizon 2020 Graphene Flagship Project (Core 3), I.V.G., A.K.G.; the Lloyd's Register Foundation, A.K.G., W.K.; A.K.G. acknowledges support from the Engineering and Physical Sciences Research Council (EPSRC; Grant Nos EP/V007033/1 and EP/Z531121/1) and ERC/UKRI-2023-ADG 2XPLO2D, grant agreement 101117904; I.V.G. acknowledges support from the EU Horizon 2020 MSCA-RISE-2019 programme (Project No. 873028 HYDROTRONICS); W.K., L.P and Z.J. acknowledge support from the EPSRC CDT in Science and Applications of Graphene and Related Nanomaterials (EPSRC Grant EP/L01548X/1). W.K. acknowledges support from National Natural Sciences Foundation of China (NSFC), Grant no. 12204546.

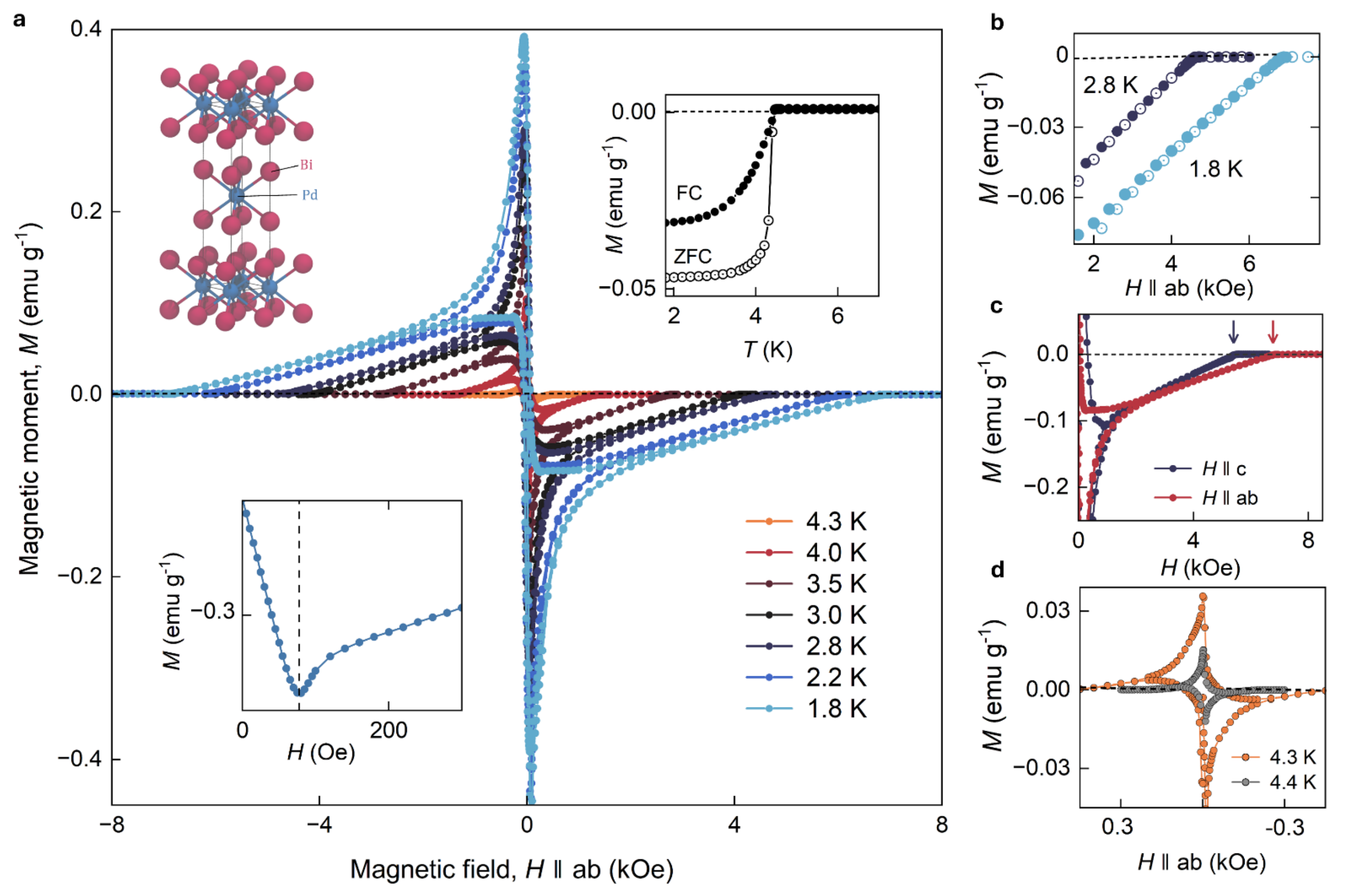

FIG. 1. **(a)** DC magnetization of β-$PdBi_2$ in magnetic fields parallel to the crystal's ab plane. *Main panel*: $M(H)$ curves at different temperatures, see legend. A sharp transition from conventional hysteretic $M(H)$ curves at low fields to strictly linear, reversible magnetization at $H > 0.3H_{c2}(T)$ is seen for all temperatures. *Top right inset:* Superconducting transition measured in field cooling (FC) and zero-field-cooling (ZFC) modes. *Top left:* Crystal structure of β-$PdBi_2$: the unit cell is composed of two Bi bilayers around central Pd atoms, with the bilayers shifted with respect to each other to form AB stacking [15]. *Bottom left*: Zoom of the magnetization curve around $H_{c1}^{\|ab}$; dashed line indicate the vortex penetration field. **(b)** High-field parts of $M(H)$ curves at $T$ = 1.8 and 2.8K emphasizing the linearity and reversibility of the magnetization and the sharp kinks at $H_{c2}$. Open symbols correspond to increasing- and solid symbols to decreasing $H$. **(c)** Comparison of the $M(H)$ curves for $H\|$ab and $H\|$c ; $T$ = 1.8 K. Notable are nearly identical values of the magnetic moment $M$ for the two field orientations and sharp kinks at $H_{c2}$ indicated by arrows. **(d)** Magnetization curves at temperatures close to $T_c$ : conventional behavior, characterized by significant hysteresis, is seen for the whole range of magnetic fields; no region of linear $M(H)$ is present at these $T$.

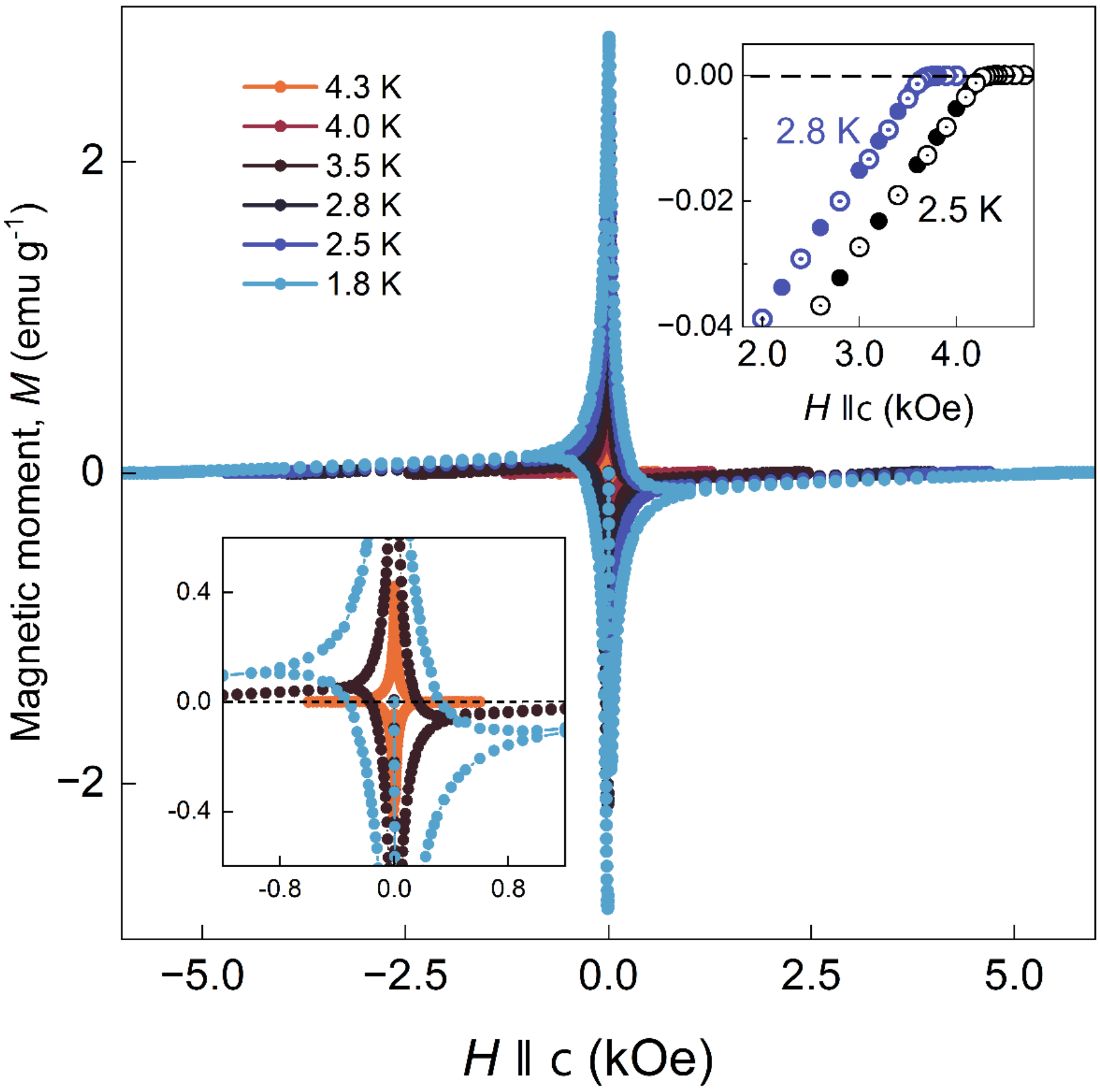

FIG. 2. DC magnetization of β-$PdBi_2$ in magnetic field perpendicular to the ab plane. *Main panel:* Magnetization curves, $M(H \| \mathrm{c})$, at different temperatures, see legend. At each temperature, the sample was cooled from $T > T_c$ in zero field; then the field swept from zero to the maximum positive, then to the maximum negative value, and back to 0. *Top right inset*: High-field parts of $M(H \| \mathrm{c})$ curves at two different temperatures (see labels) emphasizing strictly linear, reversible magnetization. Open symbols correspond to the increasing- and solid symbols to decreasing field. *Bottom left inset:* Zoom of the low-field parts of the magnetization curves (same color-coding as in the main panel) emphasizing hysteretic magnetization.

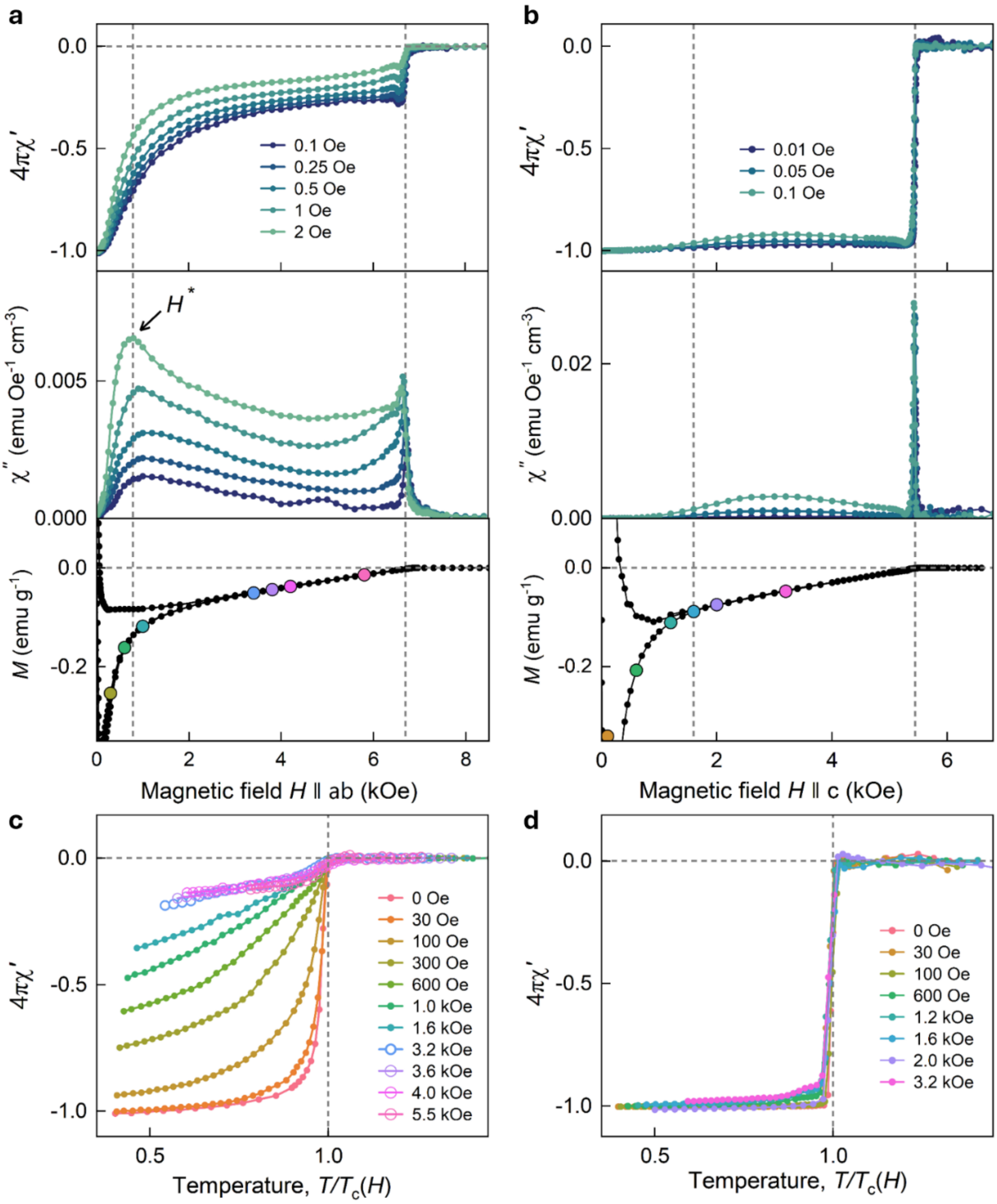

FIG. 3. AC susceptibility as a function of parallel **(a,c)** and perpendicular **(b,d)** magnetic field. **(a)** *Top and middle panels:* Real and imaginary parts of ac susceptibility, $\chi'$ and $\chi''$, measured at different excitation amplitudes $h_{ac}^{\parallel}$ (see legend) as a function of the parallel field $H||ab$ at 1.8 K. The shown dimensionless $\chi'$ corresponds to the measured volume susceptibility in units of emu·Oe$^{-1}$·cm$^{-3}$ multiplied by $4\pi$, to compare with the ideal diamagnetic response, $\chi' = -1/4\pi$. The field $H^*$, corresponding to the peak in $\chi''$ and indicated by the arrow, is taken as the transition field (see text). *Bottom panel:* Corresponding dc magnetization curve, $M(H||ab)$. Over the range of fields $H^* < H \leq H_{c2}$ where $M(H)$ is strictly linear and reversible, the surface diamagnetic screening ($\chi'$) is anomalously low, 15-25% of the ideal response, $\chi' = -1/4\pi$, and $\chi''$ indicates significant dissipation. **(b)** Same as (a) but for the perpendicular field, $H||c$. Conventional behavior $\chi' \approx const = -1/4\pi$ and $\chi'' \approx 0$ is seen over the whole range of fields up to $H_{c2}^{||c}$. **(c)** AC susceptibility, $\chi'$, as a function of reduced temperature $T/T_c(H)$ at different dc fields $H||ab$ (see legend). $h_{ac}^{\parallel} = 1$ Oe. Here $T_c(H)$ is the critical temperature corresponding to the value of $H$ for each data set. Color coding corresponds to the values of the dc field shown as large colored symbols in the bottom

panel of (a). For dc fields corresponding to linear, reversible $M(H)$ ($H^{\|ab} = 3.2$ to 5.5 kOe) all $\chi'(T)$ collapse on the same curve, saturating at low $T$ at just 15% of the full diamagnetic response that would be expected for a conventional superconductor (see text). **(d)** Same as (c) but for the $c$-axis fields. Conventional behavior, $\chi' \approx -1/4\pi$ is seen for all fields up to $H_{c2}^{\|c}$.

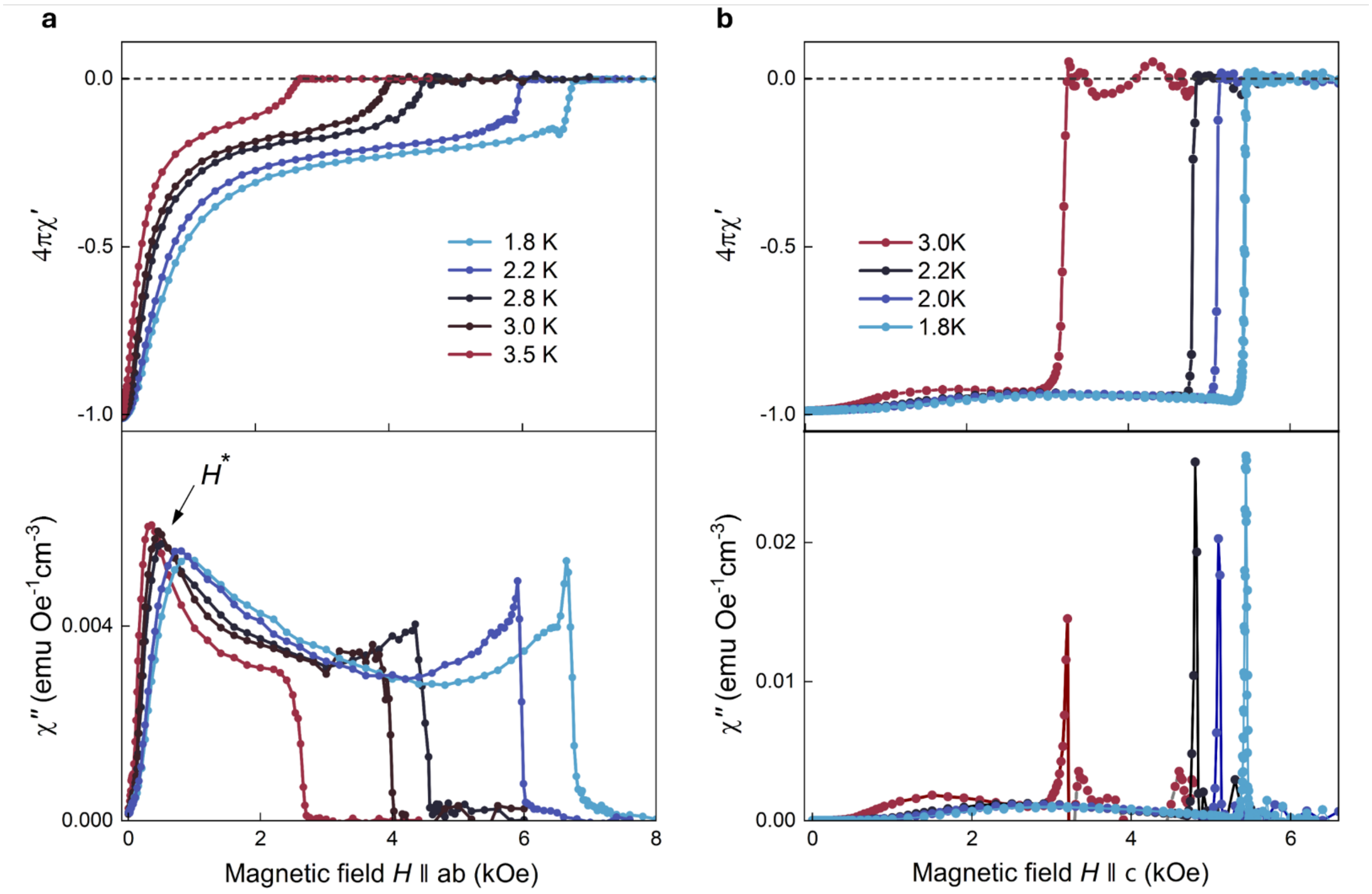

FIG. 4. AC susceptibility as a function of dc field at different temperatures. **(a)** $\chi'$ and $\chi''$ vs the in-plane magnetic field $H \| ab$. Temperatures are shown in the legend. The same qualitative behavior is seen at all $T$, i.e., a strong suppression of diamagnetic screening at $H \sim H^*$ accompanied by a peak in $\chi''$, with $H^*$ almost independent of temperature. $h_{ac}^{\|} = 1$ Oe. **(b)** Same as (a) but for the perpendicular field $H \| c$. Conventional response is seen at all $T$.

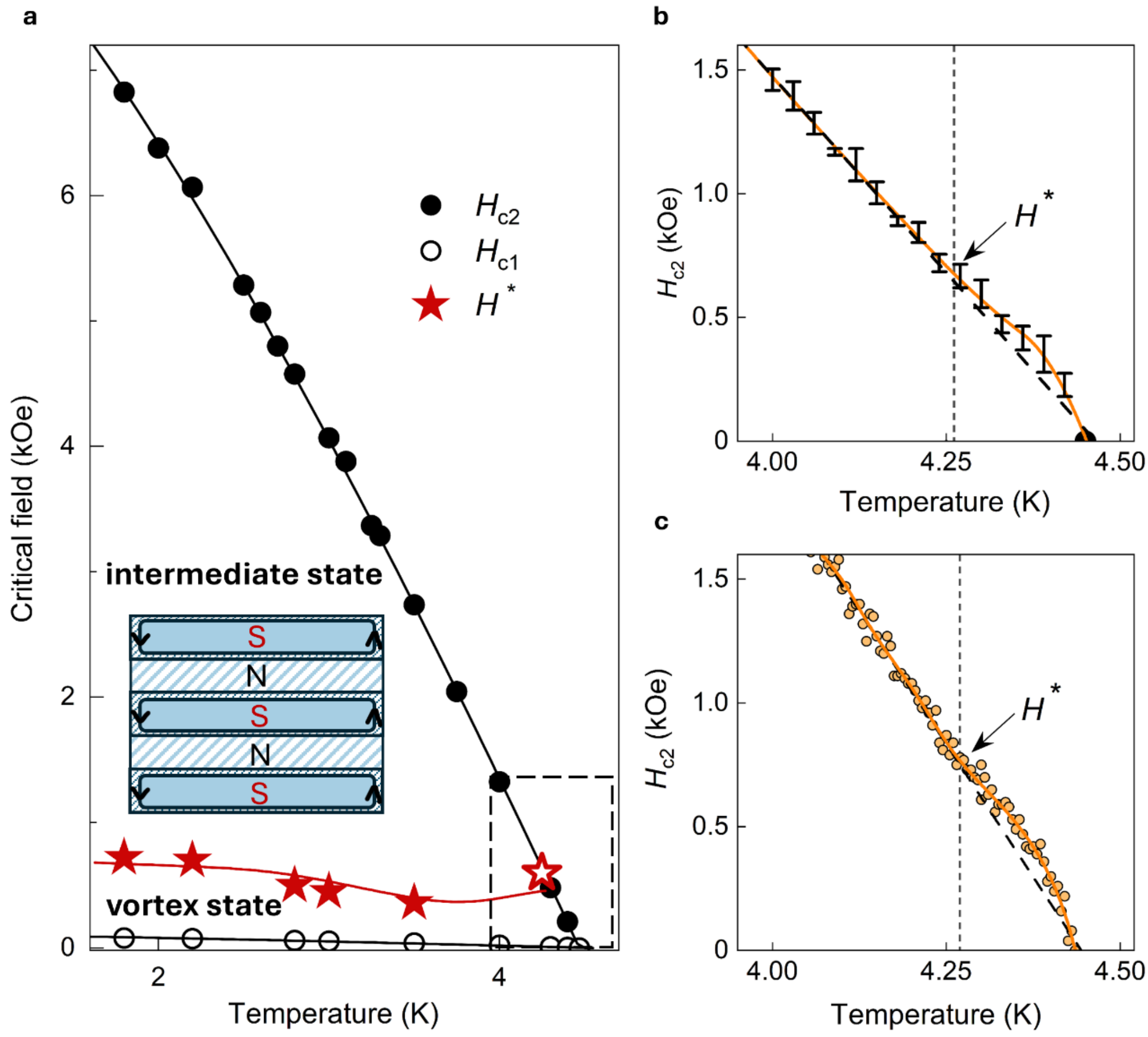

FIG. 5. Phase diagram of β-$PdBi_2$ for in-plane magnetic fields $H||ab$. **(a)** A new critical field, $H^*(T)$, is shown in addition to $H_{c1}(T)$ and $H_{c2}(T)$. Filled stars show $H^*$ values corresponding to peaks in $\chi''(H)$ at different $T$, see Fig. 4a. Open star shows the value of $H^*$ corresponding to the kink in $H_{c2}(T)$ seen in (b) and (c). Black solid lines are fits to the standard WHH theory. Red solid line is a guide to the eye. The schematic shows the proposed domain structure in the intermediate state above $H^*$. **(b)** Zoom-in of the part of the phase diagram in (a) close to $T_c$. Error bars indicate the measurement accuracy for $H_{c2}$ in this temperature range. Solid line is the guide to the eye. Dashed line emphasises the kink in $H_{c2}(T)$. **(c)** Zoom-in of the high-$T$ part of $H_{c2}(T)$ obtained from transport measurements (Fig. S3 in Supplemental Materials). A clear deviation from the linear dependence conventionally seen near $T_c$ and a kink identical to that in (b), is seen at $T\sim$ 4.25K, corresponding to $H^* \approx 0.7$ kOe.

Supplemental Material for

# Unconventional magnetization in the multiphase superconductor $PdBi_2$

Wenjun Kuang, Ziyi Jiang, Lewis Powell, Sofiia Komrakova, Andre K. Geim, Irina V. Grigorieva

## 1. Crystal growth and characterization

Single crystals of β-$PdBi_2$ were grown using a melt growth method. High-purity Pd (99.99%) and Bi (99.999%) granules with molar ratio of 1:2 were vacuum-sealed (<$10^{-4}$ mbar) in a quartz tube and melted at 1050 °C for 6 hours to ensure complete mixing of the components. The temperature was then reduced to 920 °C at 50 °C/h and maintained at this level for 24 hours. After that, the molten mixture was slowly cooled down to 500 °C at a rate of 3 °C/h and then rapidly quenched into icy water. The grown crystals were taken out from the quartz tube in the protective atmosphere of an argon filled glovebox ($O_2$<0.5 ppm, $H_2O$ < 0.5 ppm). The obtained ~1cm$^3$ crystals were easily cleavable and used to exfoliate 10 to 140μm thick crystals with clean flat surfaces that were used for magnetization measurements (details in the next section). Monocrystallinity and phase purity of the grown samples was confirmed by X-ray diffraction using Cu Kα radiation (λ = 1.5418 Å), Rigaku Smartlab. The sharp (00l) peaks with FWHM ≲ 0.03 deg (Fig. S1) correspond to a pure tetragonal β-phase of $PdBi_2$.

**Fig. S1.** Typical X-ray diffracti⁻n pattern for as-grown β-$PdBi_2$ crystals. Sharp (00l) peaks accurately match the β-phase of $PdBi_2$.

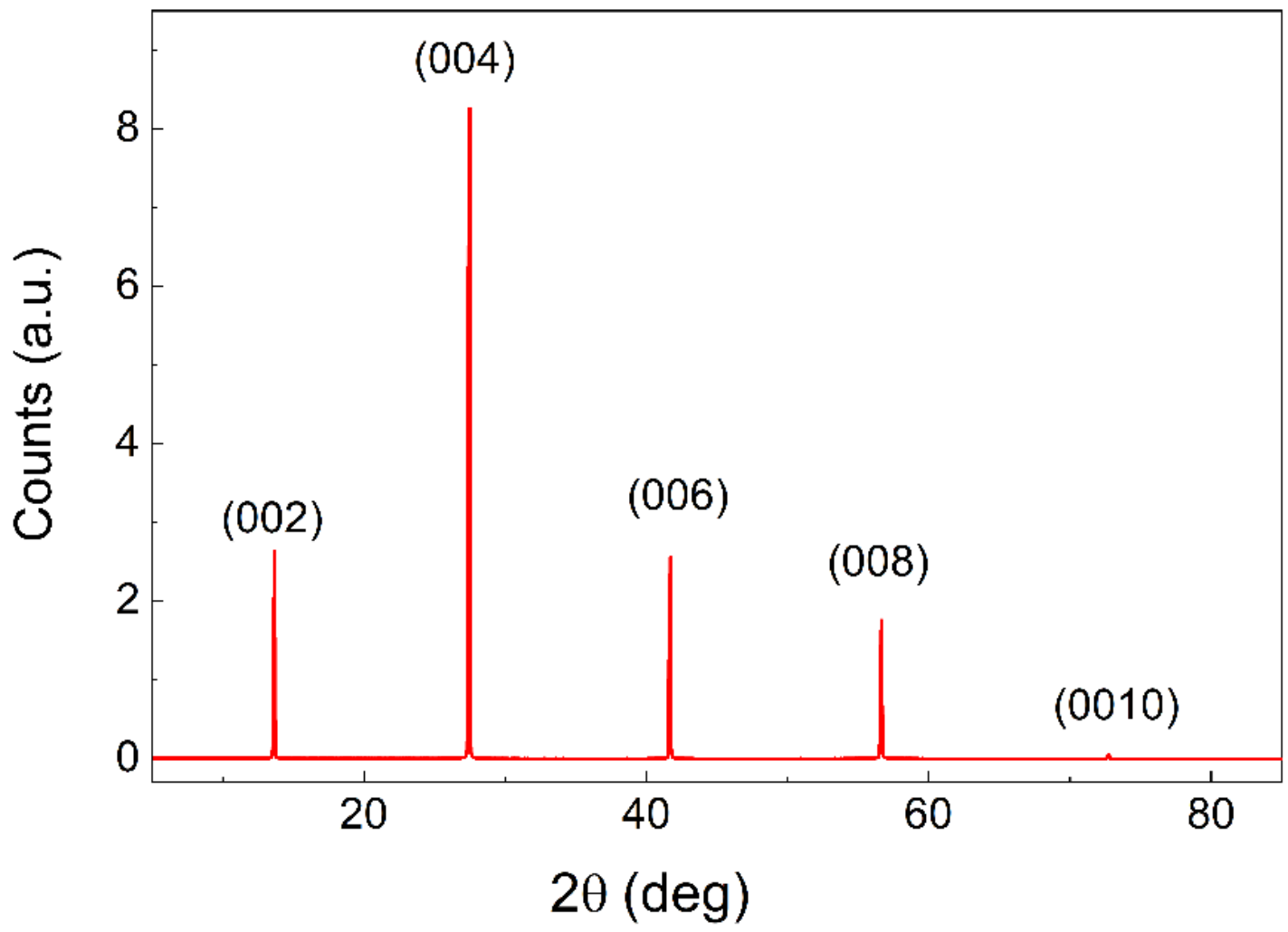

## 2. DC magnetization and ac susceptibility measurements

DC magnetization and ac susceptibility were measured using a commercial SQUID magnetometer (MPMS XL7, Quantum Design). Prior to each measurement, the surfaces of exfoliated ~15 to 140 μm thick crystals with typical lateral sizes of ~ 3 × 3 mm were cleaned by cleaving off the top layers with an adhesive tape. Before loading into the magnetometer, the samples were mounted in either parallel or perpendicular orientation with respect to the magnetic field of the magnetometer, in low magnetic background plastic straws. To align the crystals with the magnetic field ($H||$ab), they were mounted on long, stiff plastic strips which were placed tightly inside the plastic straws and arranged vertically in the narrow magnetometer insert. The stiffness of the overall arrangement prevented movement of the crystals during the measurements and ensured the accuracy of the angular alignment within $\pm$1deg. To avoid air exposure and possible degradation of the surface, $PdBi_2$ crystals were always handled in the inert atmosphere of an Ar-filled glovebox, including cleavage, exfoliation and all mounting steps.

In zero-field cooling (ZFC) mode of dc magnetization measurements, the sample was first cooled down to the lowest available temperature (1.8 K) in zero magnetic field, then a desired field applied and magnetization measured as a function of an increasing temperature $T$. In field-cooling (FC) mode, the field $H$ was applied above $T_c$ (typically at 10 to 15 K) and the magnetic moment measured as a function of decreasing temperature. All ac susceptibility data were acquired with the ac field parallel to the dc field at an excitation ac amplitude from 0.01 to 4 Oe and a frequency of 8 Hz. Test measurements of ac susceptibility at frequencies between 8 and 1400 Hz showed that the results were independent of frequency. All measurements (dc magnetization and ac susceptibility) were done in 'settle mode' (with respect to either field or temperature), where a sufficient time delay ensured that the target temperature or magnetic field was approached and stabilized before taking each data point. In addition, a given value of $H$ was approached with 'no overshoot' to avoid hysteretic effects.

The superconducting fraction $f$ was estimated from the slope of the dc $M(H)$ curves below $H_{c1}$: $f = (1 - N)4\pi|dM/dH|/V$, where $N$ is the demagnetization factor (discussed in the next section) and $V$ the sample's volume. All our measured samples yielded $f = 1$, indicating that the crystals were 100% superconducting. The measured critical fields were highly reproducible for all studied crystals and did not depend on the crystal thickness within the studied thickness range (10 to 140 μm). At the lowest measurement temperature, $T$ =1.8 K, we found $H_{c2}^{\|ab}$(1.8 K) = 6.8 kOe, $H_{c2}^{\|c}$(1.8 K) = 5.6 kOe. The temperature dependence of the upper critical field, $H_{c2}(T)$, is accurately described by the standard WHH theory [2], see Fig. 5 in the main text, yielding extrapolated zero-temperature values $H_{c2}^{\|ab}(0) \approx 9$ kOe, $H_{c2}^{\|c}(0) \approx 7.4$ kOe. The superconducting coherence length, ξ, and magnetic field penetration depth, λ, were found from the critical fields $H_{c1}$ and $H_{c2}$ using the standard expressions [1] $H_{c2} = \Phi_0/2\pi\xi^2$ and $H_{c1} = (\Phi_0/4\pi\lambda^2)[\ln\kappa + \alpha(\kappa)]$, where $\alpha(\kappa) = 0.5 + (1 + \ln 2)/(2\kappa - \sqrt{2} + 2)$, and $\kappa = \lambda/\xi$ the Ginzburg-Landau parameter. This yielded $\xi_{ab}(0) \approx 22$ nm, $\lambda(0) \approx 240$ nm and $\kappa \approx 11$, in good agreement with the values reported in the literature [3]. To estimate the penetration depth, we assumed that the vortex penetration field in Fig. 1a, $H_p^{\|ab}$(1.8K), is close to $H_{c1}^{\|ab}$ (see the corresponding discussion in the main text) yielding an extrapolated $H_{c1}^{\|ab}(0) \approx 90$ Oe.

Volume ac susceptibility $\chi'$ and $\chi''$ was calculated from ac magnetic moments $m'$ and $m''$ as $\chi' = \frac{m'}{V \cdot h_{ac}}$, $\chi'' = \frac{m''}{V \cdot h_{ac}}$, where $m'$ and $m''$ are measured in units of emu, $V$ is the sample volume in $cm^3$ and $h_{ac}$ the ac field amplitude in Oe. To compare with the ideal diamagnetic response $4\pi\chi' = -1$, $\chi'$ in Fig. 3 and 4 in the main text is shown as volume susceptibility ($emu \cdot Oe^{-1} \cdot cm^{-3}$) multiplied by $4\pi$, and the out-of-phase $\chi''$ is shown in units of $emu \cdot Oe^{-1} \cdot cm^{-3}$.

### 3. Demagnetization correction for ac susceptibility

When an external magnetic field $H_{ext}$ is applied to a superconducting film, the internal magnetic field experienced by the sample is given by (in CGS units)

$$H_{in} = H_{ext} - 4\pi NM, \tag{1}$$

where $M$ is the magnetization of the sample, and the demagnetization factor $N$ depends on the sample geometry and its orientation relative to the applied field. For the thin film geometry relevant to our study, the demagnetization factor for the perpendicular field is [3]

$$N_\perp = (1 - \gamma^2)^{-1}\left[1 - \gamma(1 - \gamma^2)^{-\frac{1}{2}}\cos^{-1}\gamma\right] \tag{2}$$

where $\gamma = d/a$ is the ratio of the thickness $d$ to the lateral size $a$ of the crystal. For $d = 15$ to ~140 μm and $a = 2-3$ mm as in our case, $N_\perp$ is in the range of 0.93-0.99, that is, close to 1, which produces a very large demagnetization field, giving a significant demagnetization effect in the superconducting state where $|M|$ is large. For measurements in parallel fields, the demagnetization factor can be neglected since $N_\parallel = (1 - N_\perp)/2 \approx 0$.

To obtain the intrinsic ac susceptibility in perpendicular fields, it is necessary to apply a demagnetization correction:

$$\chi \equiv \chi' + i\chi'' = \frac{m}{h_{\text{in}}} = \frac{m}{h_{\text{ext}} - 4\pi N_\perp m} = \frac{\chi_{\text{ext}}}{1 - 4\pi N_\perp \chi_{\text{ext}}} \tag{3}$$

where $h_{\text{in(ext)}}$ is the internal (external) ac field, $m$ the induced ac magnetization, $\chi_{\text{ext}} \equiv m/h_{\text{ext}} = \chi'_{\text{ext}} + i\chi''_{\text{ext}}$ the measured susceptibility, and $\chi$ the intrinsic ac susceptibility of the sample. Combining the above expressions for $\chi'$ and $\chi''$, one finds

$$\begin{aligned} \chi' &= \frac{\chi'_{\text{ext}} - 4\pi N_\perp \left({\chi'_{\text{ext}}}^2 + {\chi''_{\text{ext}}}^2\right)}{(4\pi N)^2\left({\chi'_{\text{ext}}}^2 + {\chi''_{\text{ext}}}^2\right) - 8\pi N_\perp \chi'_{\text{ext}} + 1} \\ \chi'' &= \frac{\chi''_{\text{ext}}}{(4\pi N)^2\left({\chi'_{\text{ext}}}^2 + {\chi''_{\text{ext}}}^2\right) - 8\pi N_\perp \chi'_{\text{ext}} + 1} \end{aligned} \tag{4}$$

Note that for the normal state, $m \approx 0$ , and $\chi = m/(h_{\text{ext}} - Nm) \approx m/h_{\text{ext}} = \chi_{\text{ext}}$ , i.e., the demagnetization effect can be neglected. Equation (4) was applied to obtain the ac susceptibility shown in Fig. 3b,d and Fig. 4b.

## 4. Amplitude dependence of the ac susceptibility in perpendicular fields

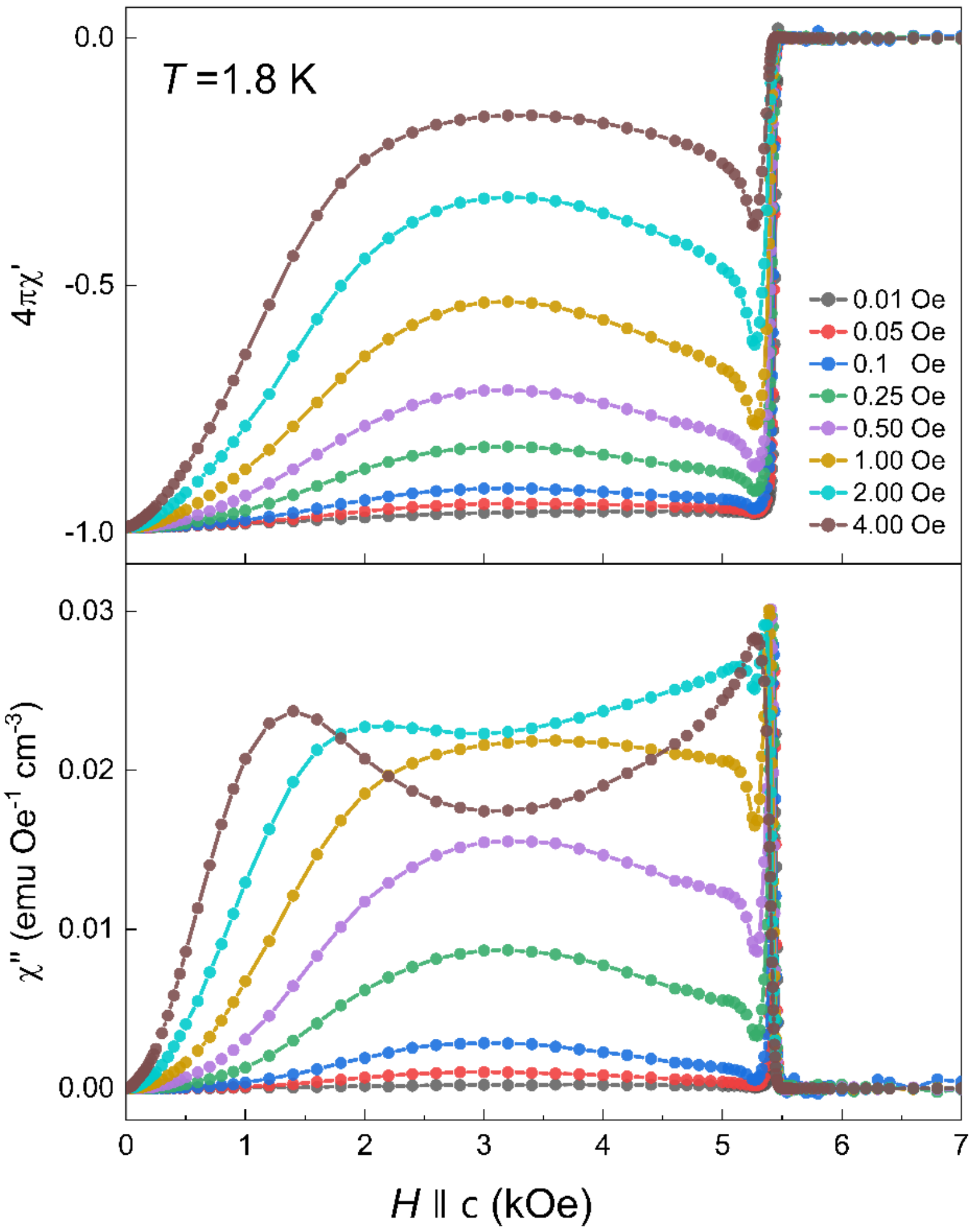

**Fig. S2. AC susceptibility in perpendicular magnetic fields at different excitation amplitudes.** At small $h_{ac} < 0.1$ Oe we observe conventional behavior, that is $\chi' = -1/4\pi$ (full diamagnetic screening) and $\chi'' \approx 0$ (no dissipation) up to the upper critical field $H_{c2}^{||c}$. At larger $h_{ac}$, the diamagnetic response is progressively suppressed, concurrent with increasing dissipation. At the largest amplitude used in our experiment, $h_{ac} = 4$ Oe, both $\chi'$ and $\chi''$ start to resemble the in-plane susceptibility, see Fig. 3 in the main text. Here it is important to note that, even though the applied field in this case is perpendicular to the crystal's ab plane, the large demagnetization factor leads to a sizable in-plane field component, such that it can exceed the relatively low value of $H^*$ and result in a domain structure similar to the case of $H||$ab. This can explain the strongly reduced diamagnetic response ($\chi'$) and large dissipation ($\chi''$) at sufficiently large amplitudes, $h_{ac} \geq 0.25$ Oe. In this case it is unfortunately not possible to separate the contribution of 'conventional' s-wave vortices from the changes due to the domain formation; this would require detailed knowledge of the domain structure and of the field penetration profile, which can be expected to be particularly complex for $H||$c due to demagnetization effects.

## 5. Transport measurements

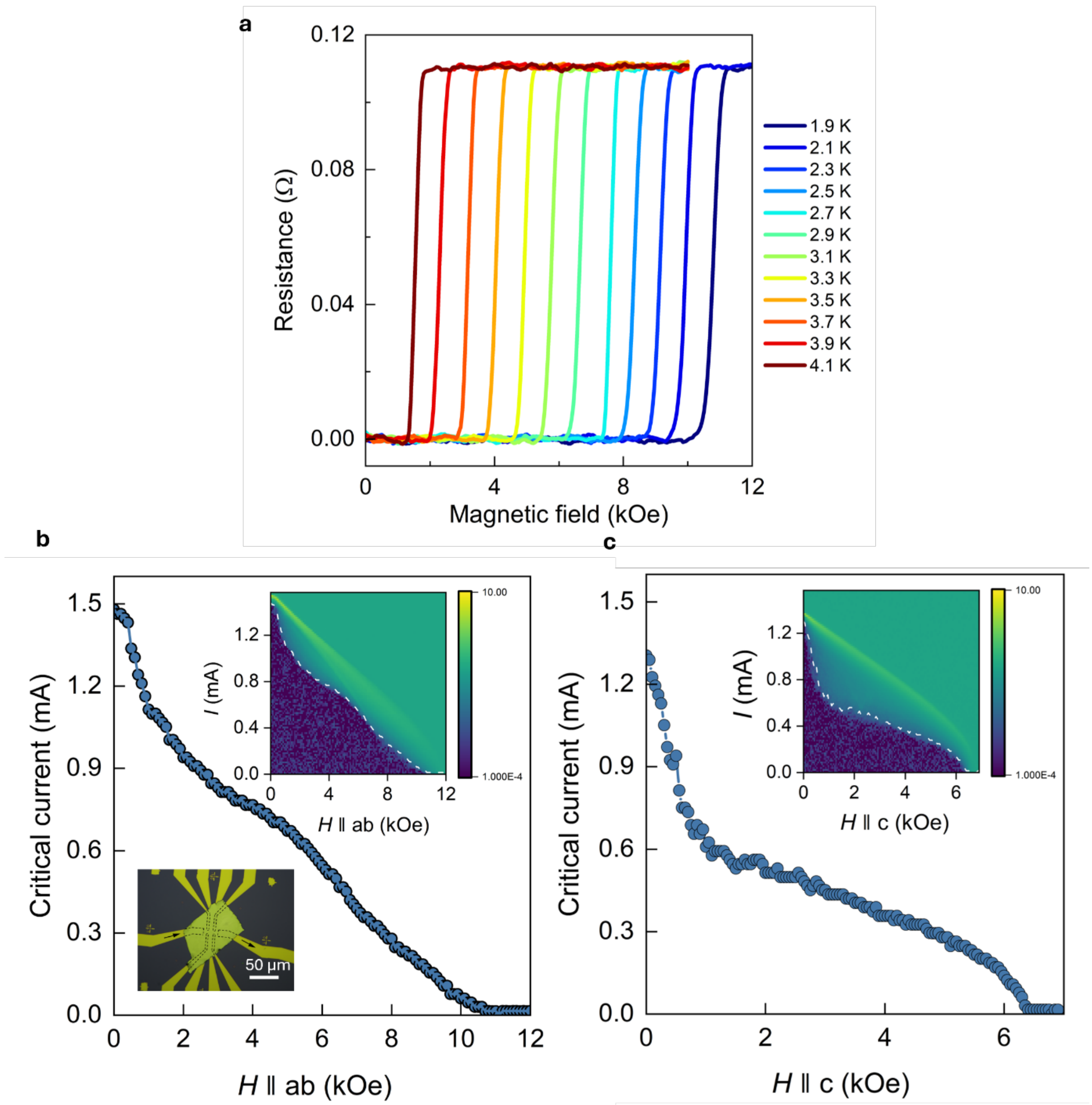

**Fig. S3. Transport measurements on a bulk β-PdBi₂ crystal.** **(a)** Resistance vs parallel magnetic field at different temperatures. **(b)** Critical current, $I_c$, measured at different magnetic fields parallel to ab plane. Temperature $T$ =1.7 K. Bottom left inset shows an optical image of the device: an exfoliated $PdBi_2$ crystal deposited onto pre-prepared Au contacts; dashed lines show the contact configuration. Assuming that the applied current flows predominantly within the 10 μm wide strip defined by the width of the current contacts, we can estimate the critical current in intermediate fields $J_c \sim 5 \cdot 10^2 \text{ A} \cdot \text{cm}^{-2}$. Top right inset shows the map of the differential resistance $dV/dI$ on a logarithmic color scale. The dashed white line corresponds to the resistance threshold used to extract $I_c$. **(c)** Same as (b) for the perpendicular field. A finite $I_c$ persists up to $H_{c2}$ for both field orientations.

As an alternative method to measure the temperature-dependent upper critical fields, $H_{c2}(T)$, and to obtain the details of the phase diagram in the low-$H$ region (Fig. 5c in the main text), we have used transport measurements, where the resistance of $PdBi_2$ crystals was measured as a function of temperature $T$, magnetic field $H$ and current $I$. These measurements also allowed us to determine the critical currents for both field orientations. To this end, ~10 μm thick crystals were exfoliated in the same way as for the magnetization measurements and placed on top of pre-patterned gold contacts on a $SiO_2$ substrate. A typical device used in these measurements is shown in the inset of Fig. S3b.

The values of $H_{c2}(T)$ were obtained from $R(H)$ measurements, such as shown in Fig. S3a, where $H_{c2}$ corresponds to $R = 0.9R_N$ and $R_N$ is the normal-state resistance. Critical currents were measured using standard lock-in techniques at $T = 1.7$K. To this end, a few mA dc current was superimposed on a $5\ \mu A$ ac current and the differential resistance, $dV/dI$, measured using a lock-in amplifier. The results are shown as insets in Fig. S3b,c. The critical current $I_c$ for each value of $H$ was defined as the current at which $dV/dI$ reached 5% of the normal-state resistance. The results for parallel and perpendicular magnetic fields are shown in the main panels of Fig. S3b,c. It is clear that for both field orientations $I_c(H)$ remains finite at all $H$, including the regions where the magnetization $M(H)$ is linear and reversible (see main text for the discussion).

### 6. Magnetization of a conventional type I superconductor

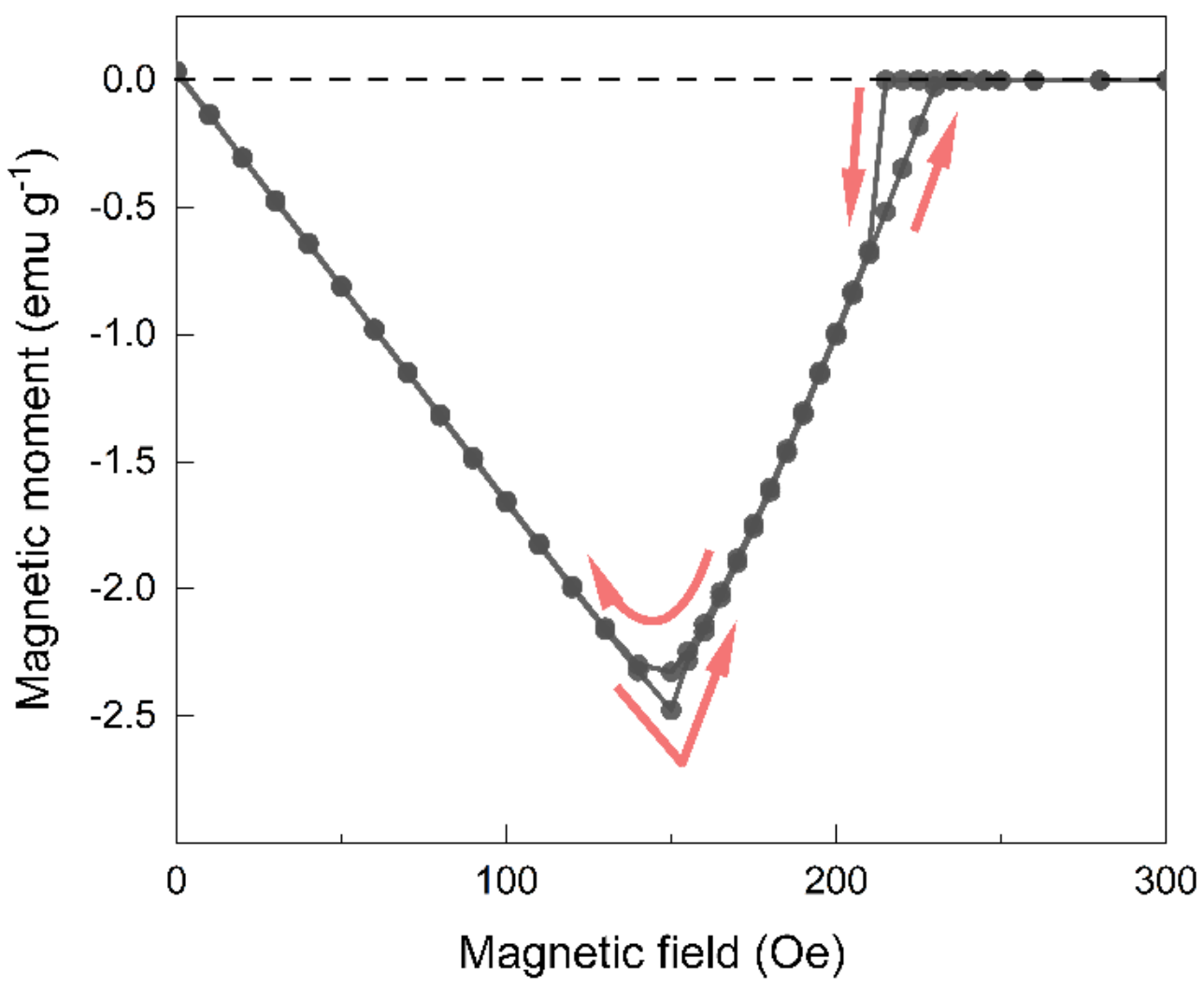

**Fig. S4. Typical dc magnetization curve for a type-I superconductor.** Shown is $M(H)$ for a cylindrical sample of Sn measured at $T$ =1.8 K; demagnetization factor $N = 0.4$. Red arrows indicate the direction of field sweep.

### References.